\documentclass[twoside,slac_one]{revtex4}
\usepackage{graphicx}
\usepackage{fancyhdr}
\usepackage{amsmath} 
\usepackage{bm}
\usepackage{amsxtra}
\usepackage{amssymb}
\usepackage{amsthm}
\usepackage{latexsym}
\usepackage{lscape}

\begin{document}

\title{A Search For The Higgs Boson In $H \rightarrow ZZ \rightarrow
  2\ell 2\nu$ Mode} 

%

\author{D. Trocino}
\affiliation{Department of Physics, Northeastern University, Boston, MA, USA}

\begin{abstract}
A search for the Standard Model Higgs boson in proton-proton
collisions at a center-of-mass energy of 7 TeV is presented in the
decay channel $H \rightarrow ZZ \rightarrow 2\ell 2\nu$. The search is 
conducted by the CMS experiment with data accumulated during the
2010 and part of the 2011 running periods of the LHC, for a total integrated
luminosity of 1.1~fb$^{-1}$. No excess is observed in the $ZZ$
transverse mass. Limits are set on the production of the Higgs boson
in the context of the Standard Model and in the presence of a
sequential fourth family of fermions with high masses. 
\end{abstract}

\maketitle

\thispagestyle{fancy}


\section{Introduction}
The search for the Higgs boson and its discovery are among the central
goals of the experiments at the LHC. Direct searches at the LEP
collider have set a lower limit on the Standard Model (SM) Higgs mass
of $114$~GeV/$c^2$~\cite{LEPHIGGS}, while direct searches at the
Tevatron 
exclude the SM Higgs in the mass range 158-173 GeV/$c^2$ at 95\%
confidence level~\cite{TEVHIGGS}. The primary production mechanism for
the Higgs at the LHC is gluon-gluon fusion, with a small
contribution from vector boson fusion~\cite{lhcxswg}. 

In the following, a search for the SM Higgs boson in the
$H \rightarrow ZZ \rightarrow 2\ell2\nu$ channel is presented~\cite{pashig11005}. This channel is
especially suited 
for high-mass Higgs boson searches, and a mass range from 250 to 600 
GeV/$c^2$ is considered in this analysis. Results are reported based on a
data sample corresponding to an integrated luminosity of
1.1~fb$^{-1}$, recorded by the CMS experiment at the LHC at a center-of-mass energy of
7~TeV. An interpretation of the results in terms of exclusion limits for
the Higgs boson is also presented in a scenario with 
a fourth generation of
fermions~\cite{cite4thgen} added to the SM. 
For sufficiently large lepton and quark masses, 
this SM extension has not been excluded by existing
constraints~\cite{tevatron4thgen}. 
Here we consider a fourth generation mass of  $600$~GeV/$c^2$.

\section{The CMS Detector}
The innermost component of the CMS detector is the tracking system,
equipped with silicon pixel and 
microstrip detectors,
used to measure the
momenta of charged particles and reconstruct the interaction vertices.
The tracker is immersed in a solenoidal magnetic field of 3.8~T and
covers the pseudorapidity range $|\eta| < 2.5$. 

The tracker is
surrounded by an electromagnetic calorimeter (ECAL), made of PbWO$_4$
crystals, and a brass/scintillating fiber hadronic calorimeter (HCAL). Both
calorimeters cover the pseudorapidity range $|\eta| < 3.0$. 
To improve the detector hermeticity, forward quartz-fiber Cherenkov
calorimeters (HF) extend the coverage up to $|\eta| = 5.0$. 

The outermost part of the CMS detector is the muon spectrometer, used
to measure the momentum of muons traversing through the detector. It
consists of three different types of gaseous detectors embedded in
the iron return yoke: drift tube chambers (DT) in the barrel, cathode
strip chambers (CSC) in the endcaps, and resistive plate chambers
(RPC) in both regions. 

A detailed description of the CMS detector can be found in
\cite{CMSDETECTOR}. 

\section{Simulation of Physics Processes}
In this analysis, signal and background datasets are produced by
detailed Monte Carlo simulation of the detector response, based on
{\sc geant4}~\cite{GEANT4,CMSDETECTOR}, taking into 
account the limited inter-calibration and alignment precision, and
using the full CMS reconstruction chain. 

\begin{figure}[ht]
  \centering
  \includegraphics[width=80mm]{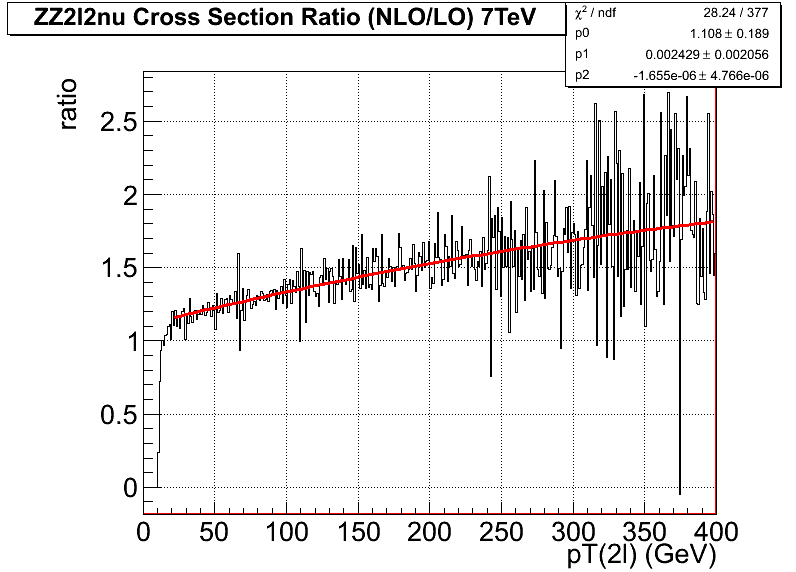}
  \hspace{5mm}
  \includegraphics[width=68mm]{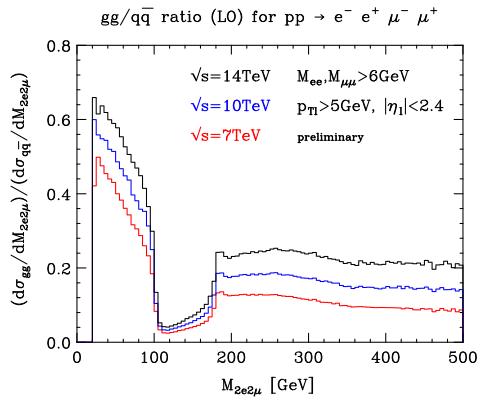}
  \caption{{\it left}: NLO to LO cross section ratio for $ZZ
    \rightarrow 2\ell2\nu$ at $\sqrt{s} = 7$~TeV as a function of the
    dilepton transverse momentum, computed using {\sc mcfm}; {\it
      right}: contribution of the $gg \rightarrow ZZ$ subprocess
    relative to $qq \rightarrow ZZ$ contribution.} 
  \label{zzHOfactors}
\end{figure}

The $H \rightarrow ZZ \rightarrow 2\ell2\nu$ signal and the Drell-Yan
and $t\bar{t}$ backgrounds are generated using the program {\sc
  powheg}~\cite{POWHEG}. The $W$~+~jets and single top events are
simulated using {\sc madgraph}~\cite{MadGraph}, while the diboson
backgrounds ($ZZ$, $WZ$, $WW$) are simulated using {\sc
  pythia}~\cite{PYTHIA}. In case of the irreducible $ZZ$ background, a
dynamic next-to-leading order $k$-factor is applied as a function
of the transverse momentum of the $Z$ boson (see
Fig.~\ref{zzHOfactors}, {\it left}). This  $k$-factor has been computed using
the {\sc mcfm} program~\cite{MCFM}. Contributions from the $gg
\rightarrow ZZ$ events are also taken into account~\cite{gluonQCD} (see 
Fig.~\ref{zzHOfactors}, {\it right}). 

\section{Event selection}
If the mass of the Higgs boson is much larger than twice the mass of
the $Z$ boson, each $Z$ boson produced in the Higgs boson decay has a
significant transverse momentum ($p_{T}$). Consequently, a $H
\rightarrow ZZ \rightarrow 2l 2\nu$ event is characterized by 
the presence of a boosted pair of isolated leptons, $e^{+}e^{-}$ or 
$\mu^{+}\mu^{-}$, with invariant mass consistent with the $Z$ mass, and
large missing transverse energy (MET) arising from the decay of the
other $Z$ boson into neutrinos. 

\subsection{Trigger and Preliminary Selections}
Since the final state involves two high-$p_{T}$ isolated leptons,
events selected for the analysis are required to pass dilepton
triggers. In case of the muon channel, a trigger with a threshold of
$p_T > 7$~GeV/$c$ for each muon was used in the initial data-taking
period (about 191 pb$^{-1}$), while for the remainder of the dataset
the thresholds were  13 and 8~GeV/$c$ for the leading and second muon,
respectively. The efficiency for selecting signal events with these
dimuon triggers is found to vary between 95 and 99\%, depending on 
the transverse momentum and pseudorapidity of the muons. In case of
the electron channel, the trigger has thresholds of $p_T > 17$~GeV/$c$
on the leading electron and $p_T > 8$~GeV/$c$ on the second
electron. The efficiency for this trigger is found to be larger than
99\%. 

In order to suppress events with an incorrect MET measurement due to
detector noise, a 
filter is applied to remove events in which anomalous HCAL noise is  
detected. Beam-halo events are also vetoed. 

An event is required to have at least one primary vertex within 24~cm
of the geometrical center of the CMS detector in the beam direction 
and within 0.2~cm in the plane transverse to the beam. In events with
multiple vertices, the one with the largest value of $\sum p_T^2$ for
the associated tracks is chosen as the reference vertex. 

\subsection{Lepton and Dilepton Selection}
Muon candidates can be reconstructed using two algorithms: one uses
tracks in the inner silicon tracker matched to segments in the muon
system; the other algorithm uses tracks produced from a global fit of
tracker and muon 
chamber hits, seeded by segments in the muon
system~\cite{CMS-PAS-MUO-10-004}. Further identification criteria
based on the number of hits in the tracker and the muon system, the
fit quality of the muon track and its consistency with the primary
vertex, are imposed on the muon candidates to reduce
fakes. 

Electron reconstruction also involves two algorithms, one in which 
energy clusters in the ECAL are matched to hits in the silicon 
tracker, and another in which tracks in the silicon tracker are 
matched to the ECAL clusters~\cite{CMS-PAS-EGM-10-004}. These
reconstructed electrons are required to pass further identification
criteria, based on the ECAL shower shape, track-ECAL
cluster matching, and consistency with the primary vertex. Additional
requirements are imposed to remove electrons produced in photon
conversions. 

To further suppress the QCD background, leptons are required to be
isolated: a cone of radius $\Delta R = \sqrt{\Delta\phi^2 +
  \Delta\eta^2} = 0.3$ is defined around the leptons, and the sum of
transverse energy deposits in the calorimeters and the transverse
momenta of tracks in this cone is computed. 
This isolation sum is required to be smaller than 15\% (10\%) of the
momentum of the muon (electron). The median energy expected from
pileup interactions 
is subtracted from the isolation sum, in order to reduce possible
inefficiencies from enhanced detector activity in events with high
pileup~\cite{FASTJET}. 

Events are selected such that there are two well-identified, isolated,
opposite-charge leptons of the same flavor, each with $p_{T} > 20$~GeV/$c$, 
which form an invariant mass consistent with the $Z$ mass. In addition,
events are required to have exactly two leptons with $p_{T} >
10$~GeV/$c$: events with any additional lepton are rejected in order 
to reduce the $WZ$ background in which the $W$ and $Z$ bosons both decay
leptonically. Moreover, the dilepton candidate is required 
to have a transverse momentum larger than 25~GeV/$c$. The impact of this
cut on signal efficiency is small (less than 1\% for all the
Higgs boson masses considered), but allows the use of $\gamma$~+~jets
data to effectively model the $Z$~+~jets background, as explained in the 
following. 

With this selection, the principal backgrounds in this
analysis are $Z$~+~jets, $t\bar{t}$, single top, $W$~+~jets, $WZ$, $WW$, and 
$ZZ$ events. 

\subsection{B-Tag Veto}
In order to suppress backgrounds containing top quarks (single top,
$t\bar{t}$), events with at least one $b$-tagged jet are
vetoed. Particle Flow jets~\cite{CMS-PAS-JME-10-003} with a
transverse energy $E_{T} > 30$~GeV are considered for $b$-tagging. 
$B$-jets are tagged
using the ``Track Counting High Efficiency'' (TCHE)
algorithm~\cite{CMS-PAS-BTV-11-001}, which uses displaced tracks in a
jet to compute a $b$-tagging discriminator. 

\subsection{Missing Transverse Energy}
The signal process $H \rightarrow ZZ \rightarrow 2\ell2\nu$ is
characterized by a large missing $E_T$, arising from the two neutrinos
in the final state. Therefore, a high threshold is imposed on the
Particle Flow MET~\cite{CMS-PAS-JME-10-005} in order to suppress the
large Drell-Yan background, which is characterized by little real
missing transverse energy. However, jet
mismeasurements and detector effects can produce large MET values in
$Z$~+~jets events. Since this background is five orders of magnitude
more abundant than the signal, the contamination of the 
selected signal region due to the high-MET tail is very
significant. To further suppress 
backgrounds with fake MET coming from jet mismeasurements, events in
which the MET is aligned with a jet are removed using a cut on the
$\Delta\phi$(MET,jet) variable. Both the MET threshold and the
$\Delta\phi$ cut are dependent on the Higgs mass hypothesis, since
high-mass Higgs events are characterized by larger MET. 

\subsection{Transverse Mass}
Higgs signal events show a narrower transverse mass distribution than
background events, where the Higgs transverse mass is defined as 
$${M_{T}}^{2} = \left(\sqrt{{p_{T,Z}}^{2} + {M_{Z}}^{2}} +
\sqrt{{\rm MET}^{2} + {M_{Z}}^{2}}\right)^{2} -
\left(\vec{p_{T,Z}} + \vec{\rm MET}\right)^{2}.$$ 
An $m_H$-dependent two-sided cut is thus applied to the $M_{T}$
variable to further separate signal from background.\\ 


The event selection criteria used in the analysis are listed in
Table~\ref{tab:selection}. For MET, $\Delta\phi$(MET,jet) and $M_T$
cuts, which are smoothly dependent on the Higgs mass, thresholds for
selected mass points are shown as examples. 

\begin{table}[htp]
  \begin{center}
    \caption{Event selection cuts.}
    \begin{tabular}{|lll|lll|} 
      \hline
      & \bf Cut &    
      & 
      & \bf Cut Value &
      \\ \hline
      & Lepton transverse momenta &
      & 
      & $p_{T} > 20 $ GeV/$c$ & 
      \\ \hline
      & $Z$ mass window &
      & 
      & $|m_{\ell\ell}-91.1876$ GeV/$c^2| \leq 15$ GeV/$c^2$ & 
      \\ \hline
      & Transverse momentum of vetoed 3$^{rd}$ lepton &
      & 
      & $p_T > 10$ GeV/$c$ & 
      \\ \hline
      & $Z$ transverse momentum &
      & 
      & $ p_{T,Z} > $ 25 GeV/$c$ &
      \\ \hline
      & $b$-tag veto (jet $p_T > 30$ GeV/$c$) &
      & 
      & TCHE discriminator $<$ 2 &
      \\ \hline
      & & 
      & 
      & $m_H = 250$ GeV/$c^2$: \hspace{0.2mm} $\Delta\phi$(MET,jet) $>$ 0.62 rad & 
      \\
      & MET-jet separation &
      & 
      & $m_H = 350$ GeV/$c^2$: \hspace{0.2mm} $\Delta\phi$(MET,jet) $>$ 0.14 rad & 
      \\
      & &
      & 
      & $m_H = 500$ GeV/$c^2$: \hspace{0.2mm} --- &
      \\ \hline
      & & 
      & 
      & $m_H = 250$ GeV/$c^2$: \hspace{0.2mm} MET $>$ 69 GeV & 
      \\
      & Missing transverse energy (MET) &
      &
      & $m_H = 350$ GeV/$c^2$: \hspace{0.2mm} MET $>$ 97 GeV &
      \\
      & &
      & 
      & $m_H = 500$ GeV/$c^2$: \hspace{0.2mm} MET $>$ 141 GeV &
      \\ \hline
      & & 
      & 
      & $m_H = 250$ GeV/$c^2$: \hspace{0.2mm} 216 GeV/$c^2$ $>$ $M_T$ $>$ 272 GeV/$c^2$ & 
      \\
      & Transverse mass ($M_T$) &
      &
      & $m_H = 350$ GeV/$c^2$: \hspace{0.2mm} 267 GeV/$c^2$ $>$ $M_T$ $>$ 386 GeV/$c^2$ & 
      \\
      & &
      & 
      & $m_H = 500$ GeV/$c^2$: \hspace{0.2mm} $M_T$ $>$ 336 GeV/$c^2$ &
      \\ \hline
    \end{tabular}
    \label{tab:selection}
    \end{center}
\end{table}

\section{Background Estimation}
All the backgrounds in this analysis can be divided into three
categories: 
\begin{itemize}
\item $Z$~+~jets with fake MET due to jet mismeasurement and detector
  effects; 
\item non-resonant backgrounds (i.e., events without a $Z$ resonance):
  $t\bar{t}$, single top, $WW$, $W$~+~jets; 
\item irreducible background: electroweak $ZZ$ pair production and fully
  leptonic decays of $WZ$ pairs. 
\end{itemize}

$ZZ$ and $WZ$ backgrounds are modeled using simulation, while the
remaining backgrounds ($Z$~+~jets and all non-resonant ones) are estimated
using data-driven methods. 

\subsection{$Z$ + jets Estimation}
\begin{figure}[ht]
  \centering
  \includegraphics[width=80mm]{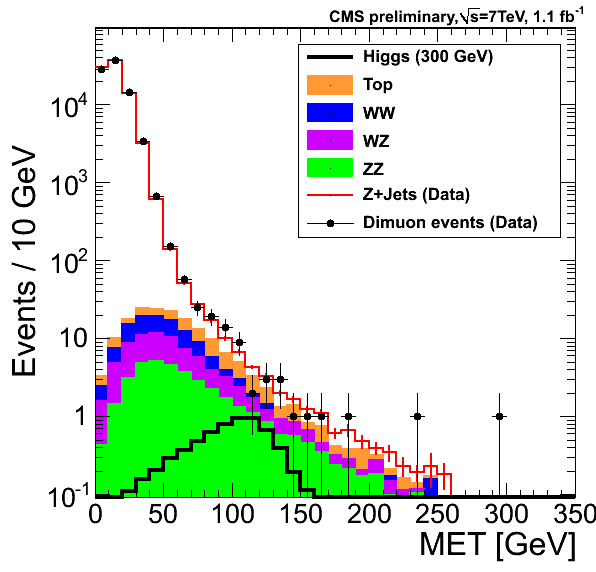}
  \includegraphics[width=80mm]{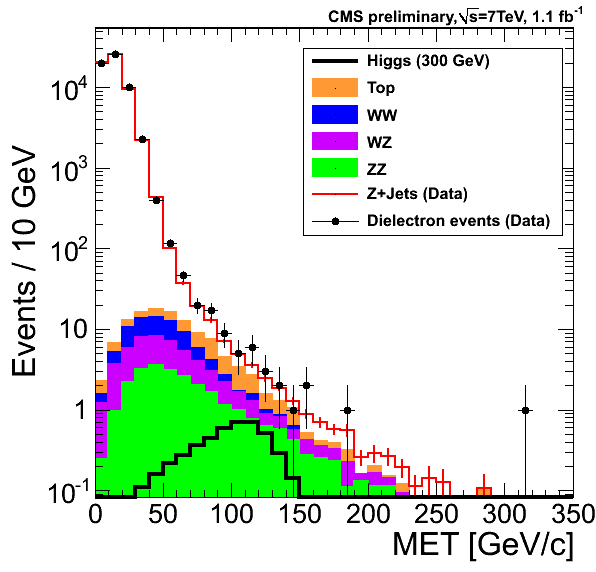}
  \caption{MET spectrum of dimuon events ({\it left}) and dielectron
    events ({\it right}), compared with reweighted single photon
    events in 1.1~fb$^{-1}$ of data.} 
  \label{fig:zgamma_met_data}
\end{figure}

The photon~+~jets process is very similar to $Z$~+~jets in terms of generating
fake MET from mismeasured jets. Moreover, the cross section for
$\gamma$~+~jets events is much greater than that of $Z$~+~jets. The former
process can thus be used to build a model of the latter, provided that
the two samples can be made kinematically identical. This is achieved
by reweighting the $\gamma$~+~jets events so that the $p_{T}$ spectrum
of the photon matches the observed dilepton $p_{T}$ spectrum. An 
additional reweighting is applied to match the jet multiplicity
between the two samples. The photon is also assigned a random mass,
according to the $Z$ lineshape observed in dilepton events, and the
yield of the photon sample is normalized to that of the $Z$. Lastly, the
full analysis selection is applied to the reweighted photon
sample. The MET distribution obtained with this procedure models very
accurately the spectrum observed in Drell-Yan events (see
Fig.~\ref{fig:zgamma_met_data}). The statistical error for the
$Z$~+~jets prediction thus obtained is computed by summing in quadrature
the weights of the $\gamma$~+~jets events. 

\subsection{Non-Resonant Background Estimation}
The non-resonant backgrounds consist of several contributions.
Events with
$t\bar{t}$ and $WW$ are characterized by a final state with two prompt
leptons from $W$ decays and real MET, plus two $b$-jets for
$t\bar{t}$. Single top events produced in the most common t$W$-channel
show a similar signature. $W$~+~jets has instead one prompt lepton and a
jet faking a second lepton. 

All non-resonant backgrounds can be estimated using events with a final
state containing $e^{+}\mu^{-}$ or $e^{-}\mu^{+}$ pairs and passing
the full analysis selection. The background level in the $e^{+}e^{-}$
and $\mu^{+}\mu^{-}$ final states is computed using 
the relations $N_{\mu\mu} = \alpha_{\mu} \times N_{e\mu}$ and $N_{ee}
= \alpha_{e} \times N_{e\mu}$, where the scale factors $\alpha$ can be
computed from the sidebands (SB) of the $Z$ peak: 
$$\alpha_{\mu} = \frac{N_{\mu\mu}^{SB}}{N_{e\mu}^{SB}}, \;\;\;\;\;\;\;\; 
\alpha_{e} = \frac{N_{ee}^{SB}}{N_{e\mu}^{SB}}.$$

This method cannot distinguish between the non-resonant background and
the signal in the $H \rightarrow WW \rightarrow 2l2\nu$ channel. Such
signal events are therefore treated as a part of the background
estimate. This signal contamination is found to be less than 7\% of
the total background yield for all considered Higgs masses. In the SM 
scenario with a fourth generation of fermions, this contamination
increases, up to about 50\% of the total background for a Higgs mass
of 250 GeV/$c^{2}$. Table~\ref{tab:R_muele_results} lists the
predicted yields for non-resonant backgrounds with 1.1~fb$^{-1}$. 

\begin{table}[htp]
  \begin{center}
    \caption{Yields from 1.1~fb$^{-1}$ data for non-resonant
      backgrounds in the muon ({\it left}) and electron ({\it right})
      channels. Statistical uncertainties are also quoted.} 
    \begin{tabular}{|lll|ccc|ccc|} \hline
      & $\mathbf m_{\mathbf H}$  & & & \bf Predicted Yields & & & \bf
      MC Prediction  & 
      \\ \hline
      & $250$  & & & $13\pm3$  & & & $12\pm0.65$  &
      \\ \hline
      & $300$  & & & $3\pm1.4$  & & & $4.8\pm0.39$  &
      \\ \hline
      & $350$  & & & $2\pm1.2$  & & & $1.4\pm0.21$  &
      \\ \hline
      & $400$  & & & $0.68\pm0.69$  & & & $0.42\pm0.11$  &
      \\ \hline
      & $500$  & & & $0$  & & & $0.038\pm0.033$  &
      \\ \hline
      & $600$ & &  & $0$  & & & $0$  &
      \\ \hline
    \end{tabular}
  \hspace{5mm}
    \begin{tabular}{|lll|ccc|ccc|} \hline
      & $\mathbf m_{\mathbf H}$  & & & \bf Predicted Yields  & & & \bf
      MC Prediction & 
      \\ \hline 
      & $250$ & &  & $9.4\pm2.2$  & & & $9.5\pm0.61$ &
      \\ \hline
      & $300$  & & & $2.2\pm1$  & & & $3.7\pm0.36$  &
      \\ \hline
      & $350$  & & & $1.3\pm0.78$  & & & $0.99\pm0.16$  &
      \\ \hline
      & $400$  & & & $0.5\pm0.5$  & & & $0.36\pm0.1$  &
      \\ \hline
      & $500$  & & & $0$  & & & $0.17\pm0.075$ &
      \\ \hline
      & $600$  & & & $0$  & & & $0$ &
      \\ \hline
    \end{tabular}
    \label{tab:R_muele_results}
  \end{center}
\end{table}

\section{Systematic Uncertainties}
The signal and background yields are subject to several systematic
uncertainties, summarized in Table~\ref{tab:systematics}. Most of
these uncertainties apply to processes derived from simulation (Higgs
signal, $ZZ$ and $WZ$ backgrounds) or to data-driven estimates of
backgrounds ($Z$~+~jets and non-resonant backgrounds.)

\begin{table}[htp]
  \begin{center}
    \caption{Summary of the main systematic uncertainties affecting
      signal and background yields.}  
    \begin{tabular}{|lll|ccc|} 
      \hline
      & \bf Source of uncertainty  & & & \bf Uncertainty [\%] & 
      \\ \hline
      & Luminosity                                     & & & 6 & 
      \\ \hline
      & PDF, gluon-gluon initial state                 & & & 6-11 & 
      \\ \hline
      & PDF, quark-quark initial state                 & & & 3.3-7.6 & 
      \\ \hline
      & QCD scale, gluon-gluon initial state           & & & 7.6-11
      ($ggH$), 20 ($ggZZ$) & 
      \\ \hline
      & QCD scale, quark-quark initial state           & & & 0.2-2
      (VBF), 5.8-8.5 ($qqVV$) & 
      \\ \hline
      & Anti $b$-tagging                                 & & & 1-1.2 & 
      \\ \hline
      & Lepton ID, isolation                           & & & 2 & 
      \\ \hline
      & Lepton momentum scale                          & & & 2
      (2$\mu$), 5 (2e) & 
      \\ \hline
      & Jet energy scale                               & & & 1.5 & 
      \\ \hline
      & Pileup                                         & & & 1-3 & 
      \\ \hline
      & Trigger                                        & & & 2
      (2$\mu$), 1 (2e) & 
      \\ \hline
      & Non-resonant background estimation             & & & 15-100 & 
      \\ \hline
      & Drell-Yan background estimation                & & & 51-54 & 
      \\ \hline
    \end{tabular}
    \label{tab:systematics}
  \end{center}
\end{table}

\section{Results and Conclusions}
\begin{figure}[h]
  \includegraphics[width=80mm]{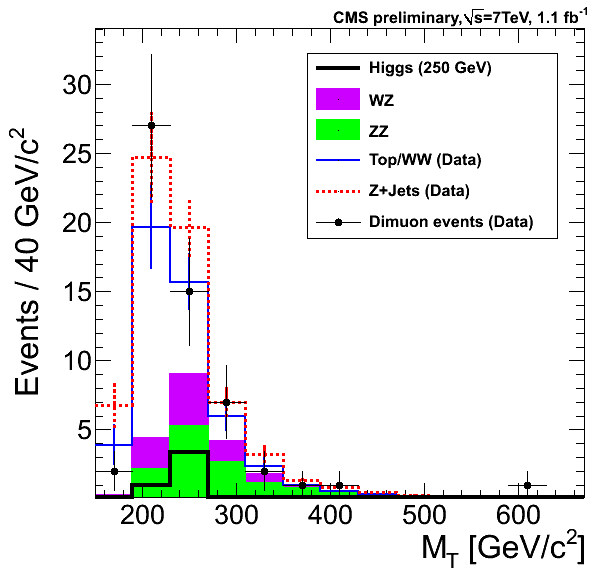}
  \includegraphics[width=80mm]{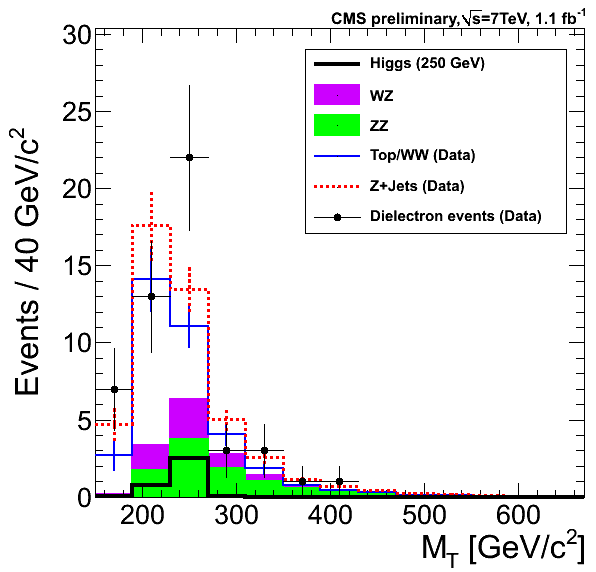}\\
  \includegraphics[width=80mm]{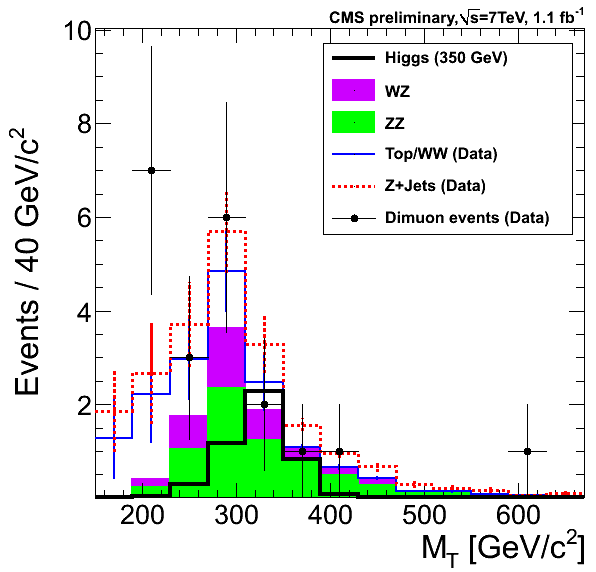}
  \includegraphics[width=80mm]{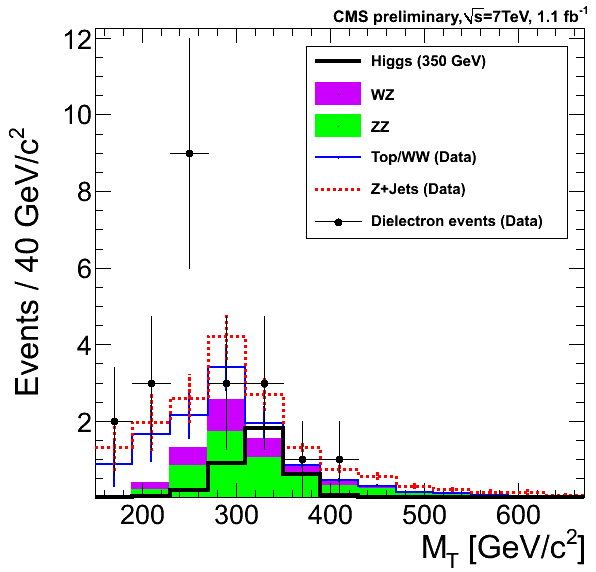}
  \caption{$M_{T}$ distribution of events passing the 250~GeV/$c^{2}$
    ({\it above}) and 350~GeV/$c^{2}$ ({\it below}) Higgs mass
    selections, in the muon ({\it left}) and electron ({\it right})
    channels.} 
  \label{fig:mt_fullsel_250}
\end{figure}

Figure~\ref{fig:mt_fullsel_250} shows the $M_{T}$ distributions of the
selected dilepton events from data and the estimated background for the
250 GeV/$c^2$ and 350 GeV/$c^2$ Higgs mass hypotheses. No evidence of
Higgs boson production is found. Table~\ref{tab:final_yields} lists
the event yields after the full selection for various Higgs mass
hypotheses, for the estimated backgrounds, and for the 1.1~fb$^{-1}$
dataset. 

\begin{table}[ht]
  \begin{center}
    \caption{Expected number of signal and background events for an 
      integrated luminosity of 1.1~fb$^{-1}$ after full analysis
      selection, for dimuon ({\it above}) and dielectron ({\it below})
      channels. Uncertainties for $ZZ$/$WZ$ include the statistical and 
      systematic components (in this order), non-resonant and $Z$+jets 
      backgrounds have the full uncertainty quoted, and, finally, the 
      signal column has only the systematic uncertainty quoted as the 
      statistical uncertainty is negligible.} 
    \begin{tabular}{|ccc|c|c|c|c|c|c|c|c|} 
      \hline
      & $\mathbf m_{\mathbf H}$ & & \bf ZZ & \bf WZ & \bf Non-resonant
      & \bf Z + Jets & \bf Total & \bf Signal & \bf Data 
      \\ \hline 
      & 250 GeV/$c^2$ & & $6.3\pm 0.1\pm 0.5$ & $4.5\pm 0.2\pm 0.5$ &
      $13\pm 3$ & $6.3\pm 3.3$ & $30\pm 5$ & $3.9\pm 0.5$ & $28$ 
      \\ \hline
      & 300 GeV/$c^2$ & & $5\pm 0.1\pm 0.4$ & $3.1\pm 0.2\pm 0.3$ &
      $3\pm 1.4$ & $3.5\pm 1.8$ & $15\pm 2$ & $3.8\pm 0.5$ & $13$ 
      \\ \hline
      & 350 GeV/$c^2$ & & $5.2\pm 0.1\pm 0.5$ & $2.5\pm 0.1\pm 0.3$ &
      $2\pm 1.2$ & $3.5\pm 1.8$ & $13\pm 2$ & $3.9\pm 0.6$ & $11$ 
      \\ \hline
      & 400 GeV/$c^2$ & & $3.6\pm 0.1\pm 0.3$ & $1.4\pm 0.1\pm 0.2$ &
      $0.68\pm 0.68$ & $2.8\pm 1.5$ & $8.6\pm 1.8$ & $3.2\pm 0.4$ & $6$ 
      \\ \hline
      & 500 GeV/$c^2$ & & $2\pm 0.9\pm 0.2$ & $0.69\pm 0.08\pm 0.08$ &
      $0$ & $1.6\pm 0.9$ & $4.4\pm 1$ & $1.4\pm 0.2$ & $3$ 
      \\ \hline
      & 600 GeV/$c^2$ & & $1.1\pm 0.1\pm 0.1$ & $0.34\pm 0.05\pm 0.04$
      & $0$ & $1\pm 0.5$ & $2.5\pm 0.7$ & $0.57\pm 0.08$ & $2$ 
      \\ \hline
    \end{tabular}

    \begin{tabular}{|ccc|c|c|c|c|c|c|c|c|} 
      \hline
      & $\mathbf m_{\mathbf H}$ & & \bf ZZ & \bf WZ & \bf Non-resonant
      & \bf Z + Jets & \bf Total & \bf Signal & \bf Data 
      \\ \hline 
      & 250 GeV/$c^2$ & & $4.6\pm 0.1\pm 0.5$ & $3.6\pm 0.2\pm 0.4$ &
      $9.4\pm 2.2$ & $4\pm 2.1$ & $22\pm 5$ & $3.1\pm 0.4$ & $29$ 
      \\ \hline 
      & 300 GeV/$c^2$ & & $3.8\pm 0.1\pm 0.4$ & $2.2\pm 0.1\pm 0.3$ &
      $2.2\pm 1$ & $2.6\pm 1.3$ & $11\pm 3$ & $3.1\pm 0.4$ & $21$ 
      \\ \hline 
      & 350 GeV/$c^2$ & & $4.2\pm 0.1\pm 0.4$ & $1.8\pm 0.1\pm 0.2$ &
      $1.3\pm 0.8$ & $3.2\pm 1.7$ & $11\pm 3$ & $3.2\pm 0.5$ & $8$ 
      \\ \hline 
      & 400 GeV/$c^2$ & & $3\pm 0.1\pm 0.3$ & $1.2\pm 0.1\pm 0.1$ &
      $0.5\pm 0.5$ & $2.6\pm 1.3$ & $7.3\pm 2$ & $2.5\pm 0.3$ & $7$ 
      \\ \hline 
      & 500 GeV/$c^2$ & & $1.6\pm 0.1\pm 0.2$ & $0.62\pm 0.07\pm 0.07$
      & $0$ & $1.7\pm 0.9$ & $3.9\pm 1.2$ & $1.1\pm 0.2$ & $2$ 
      \\ \hline 
      & 600 GeV/$c^2$ & & $0.97\pm 0.06\pm 0.09$ & $0.31\pm 0.05\pm
      0.04$ & $0$ & $1\pm 0.6$ & $2.3\pm 0.7$ & $0.47\pm 0.07$ & $1$ 
      \\ \hline 
    \end{tabular}
    \label{tab:final_yields}
  \end{center}
\end{table}

\begin{figure}[h]
  \includegraphics[width=84mm]{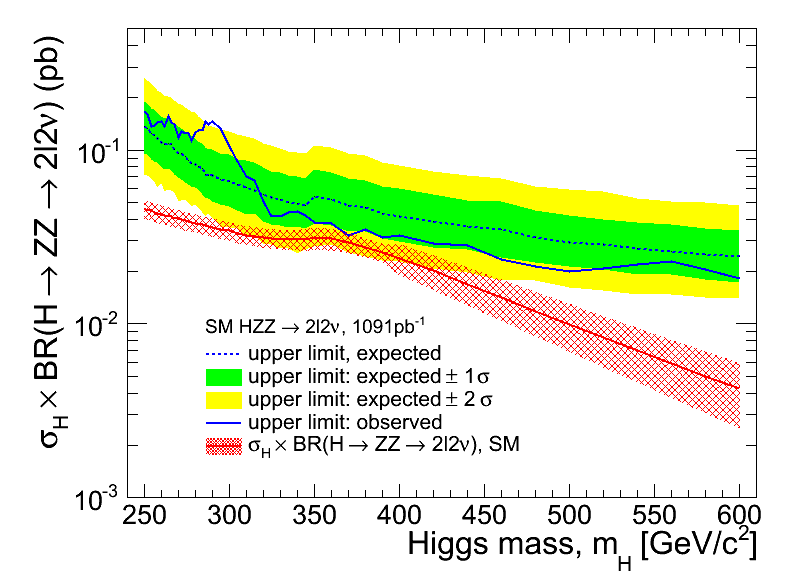}
  \includegraphics[width=84mm]{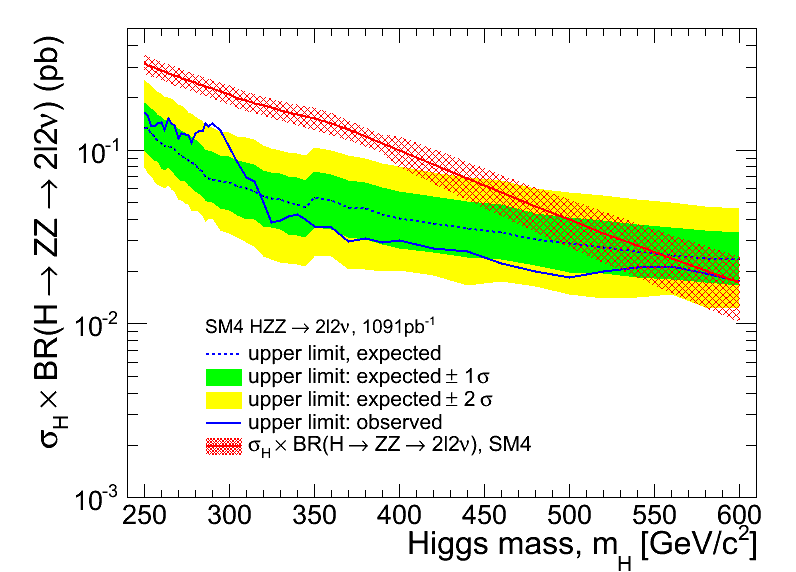}\\
  \includegraphics[width=84mm]{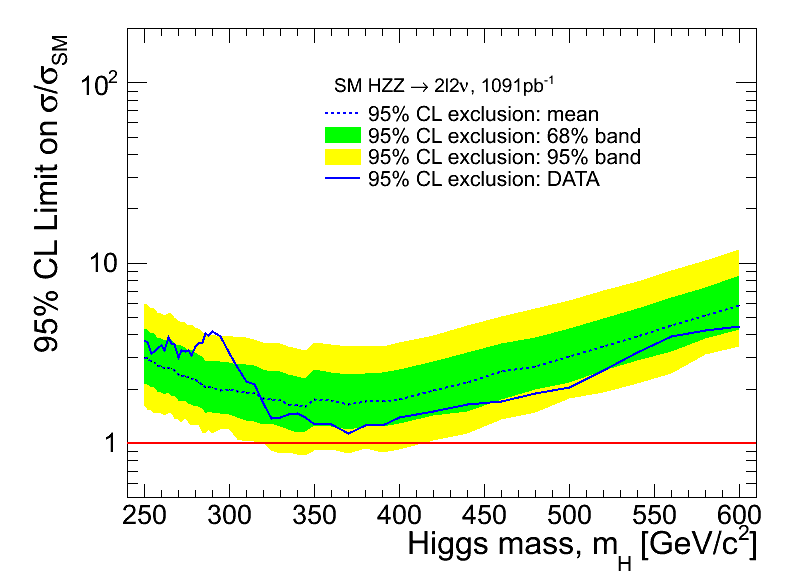}
  \includegraphics[width=84mm]{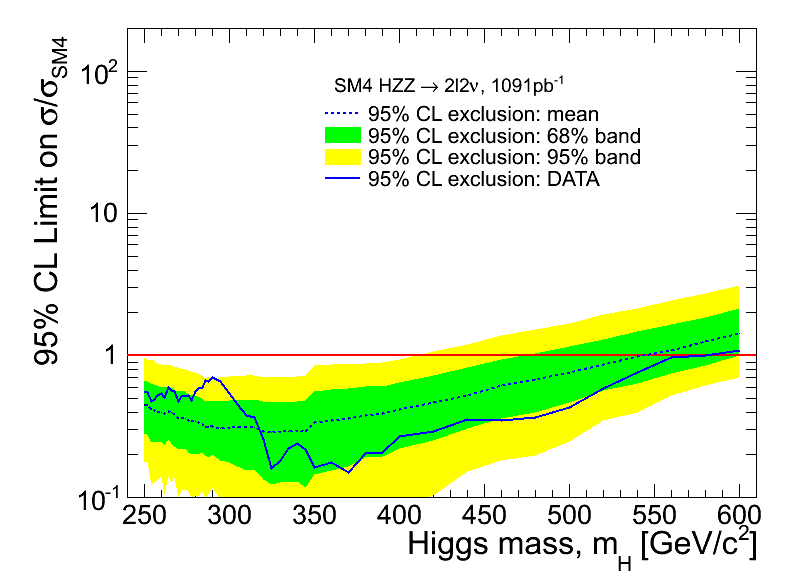}
  \caption{{\it Above:} 95\% mean expected and observed C.L. upper
    limits on the $H \rightarrow ZZ \rightarrow 2\ell2\nu$ cross
    section (in red), in the SM ({\it left}) and SM + 4$^{th}$
    generation ({\it right}). {\it Below:} ratio of the 95\% C.L. upper
    limit ($\sigma$) to the SM cross section $\sigma_{\mathrm{SM}}$
    ({\it left}), and SM + 4$^{th}$ generation cross section
    $\sigma_{\mathrm{SM4}}$. All results are relative to 1.1~fb$^{-1}$
    of integrated luminosity.} 
  \label{fig:limits}
\end{figure}

Figure~\ref{fig:limits} ({\it left}) shows the 95\% mean expected and
observed C.L. upper limits on $\sigma \times BR(H \rightarrow ZZ
\rightarrow 2\ell2\nu)$ ({\it above}), and the ratio $R$ 
of such limit to the SM cross section $\sigma_{\mathrm{SM}}$
({\it below}), as functions of the Higgs mass
$m_H$. The results are obtained using a CL$_{s}$ approach with
a flat prior for the cross section, for 1.1~fb$^{-1}$. 
Such limits are not sufficient to exclude the existence of the SM
Higgs boson in the examined mass range. 
The same results for the SM augmented with a fourth generation of
fermions with masses of $600$~GeV/$c^2$ are shown in
Fig.~\ref{fig:limits} ({\it right}). The presence of
another fermion family produces an enhancement of the dominant
gluon-gluon fusion cross section. With 1.1~fb$^{-1}$, Higgs masses in
the range $250 - 560$ GeV/$c^{2}$ can be excluded in this model. 

In this study, the exclusion limit is compared to the cross
section for on-shell Higgs production and subsequent decay in the
zero-width 
approximation. Acceptance estimates are obtained with Monte Carlo
simulations based on ad-hoc Breit-Wigner distributions for
describing the Higgs boson propagation. Recent analyses show that the
use of a QFT-consistent Higgs propagator, and allowing for the
off-shellness of the Higgs boson, dynamical QCD scales, and
interference effects between Higgs signal and backgrounds, will result,
at Higgs masses above 300~GeV/$c^{2}$, in a sizable effect on
conventionally defined but theoretically consistent parameters (mass
and width) that describe the propagation of an unstable
Higgs boson~\cite{lhcxswg,anastasiou,pasturmucci}. In this work, these
effects are included as an additional uncertainty of 10-30\% (for 400
to 600~GeV/$c^{2}$ Higgs mass) on the theoretical cross section. 

\begin{acknowledgments}
The author would like to thank the CMS Higgs PAG conveners, professors
Vivek Sharma and Christoph Paus; the conveners of the HZZ subgroup,
Dr. Chiara Mariotti and Prof. Ivica Puljak; and the whole HZZ2l2nu
working group, especially Adish Vartak, Dr. Alexey Drozdetskiy and
Dr. Petra Merkel, for coordinating the huge effort that has led to the
results here presented. 
\end{acknowledgments}

\bigskip 

\begin{thebibliography}{99} 

\bibitem{LEPHIGGS} ALEPH, DELPHI, L3, OPAL Collaborations, LEP Working
  Group for Higgs boson searches, ``Search for the Standard Model
  Higgs boson at LEP'', {\it Phys. Lett. B} {\bf 505} (2003) 61. 
\bibitem{TEVHIGGS} CDF, D0 Collaborations, TEVNPHWG Working Group,
  ``Combined CDF and D0 Upper Limits on Standard Model Higgs Boson
  Production with up to 8.2 fb-1 of Data'',
  \href{http://arxiv.org/abs/1103.3233v2}{arXiv:1103.3233v2}. 
\bibitem{pashig11005} CMS Collaboration, ``Search for the Higgs boson
  in the $H \rightarrow ZZ \rightarrow 2l2\nu$ channel in pp
  collisions at $\sqrt{s}$ = 7~TeV'', {\bf CMS-PAS-HIG-11-005}
  (2011). 
\bibitem{lhcxswg} LHC Higgs Cross Section Working Group, S. Dittmaier,
  C. Mariotti {\it et al.}, ``Handbook of LHC Higgs Cross Sections:
  1. Inclusive Observables'', {\it CERN-2011-002} (CERN, Geneva, 2011)
  {\tt arXiv:1101.0593}. 
\bibitem{cite4thgen} Q. Li, M. Spira, J. Gao, C.S. Li, ``Higgs boson
  production via gluon fusion in the standard model with four
  generations'', {\tt arXiv:1011.4484} (2010). 
\bibitem{tevatron4thgen} T. Aaltonen {\it et al.}, ``Combined Tevatron
  upper limit on $gg \rightarrow H \rightarrow W^{+}W^{-}$ and
  constraints on the Higgs boson mass in fourth-generation fermion
  models'', {\it Phys. Rev. D} {\bf 82} (2010) 011102. 
\bibitem{CMSDETECTOR} CMS Collaboration, ``The CMS experiment at the
  CERN LHC'', {\it JINST} {\bf 3} (2008) S08004. 
\bibitem{GEANT4} S. Agostinelli {\it et al.}, ``Geant 4 A
  Simulation Toolkit'', {\it Nucl. Inst. Meth. A} {\bf 506} (2003)
  250. 
\bibitem{POWHEG} S. Frixione, P. Nason, C. Oleari, ``Matching NLO
  QCD computations with Parton Shower simulations: the POWHEG
  method'', {\it JHEP} {\bf 11} (2007) 070, {\tt arXiv:0709.2092}. 
\bibitem{MadGraph} F. Maltoni, T. Stelzer, ``MadEvent: Automatic
  event generation with MadGraph'', {\it JHEP} {\bf 02} (2003), {\tt 
    arXiv:0208156}. 
\bibitem{PYTHIA} T. Sj{\" o}strand, S. Mrenna, P. Skands, ``PYTHIA
  6.4 physics and manual'', {\it JHEP} {\bf 05} (2007).
\bibitem{MCFM} J. M. Campbell, R. K. Ellis, ``MCFM for the Tevatron
  and the LHC'', {\tt arXiv:1007.3492}. 
\bibitem{gluonQCD} T. Binoth, N. Kauer, P. Mertsch, ``Gluon-induced
  QCD corrections to pp $\rightarrow$ ZZ $\rightarrow$ l anti-l
  l$^{\prime}$ anti-l$^{\prime}$'', {\tt arXiv:0807.0024}. 
\bibitem{CMS-PAS-MUO-10-004} CMS Collaboration, ``Performance of muon
  identification in 2010 data'', {\bf CMS-PAS-MUO-10-004} (2010). 
\bibitem{CMS-PAS-EGM-10-004} CMS Collaboration, ``Electron
  reconstruction and identication at $\sqrt{s}$ = 7~TeV'', {\bf
    CMS-PAS-EGM-10-004} (2010). 
\bibitem{FASTJET} M. Cacciari, G. P. Salam, G. Soyez, ``The Catchment
  Area of Jets'', {\it JHEP} {\bf 04} (2008) 005, {\tt
    arXiv:0802.1188}. 
\bibitem{CMS-PAS-JME-10-003} CMS Collaboration, ``Jet Performance in
  pp Collisions at $\sqrt{s}$ = 7~TeV'', {\bf CMS-PAS-JME-10-003}
  (2010). 
\bibitem{CMS-PAS-BTV-11-001} CMS Collaboration, ``Performance of b-jet
  identification in CMS'', {\bf CMS-PAS-BTV-11-001} (2011). 
\bibitem{CMS-PAS-JME-10-005} CMS Collaboration, ``MET Performance in
  Events Containing Electroweak Bosons from pp Collisions at
  $\sqrt{s}$ = 7~TeV'', {\bf CMS-PAS-JME-10-005} (2010). 
\bibitem{anastasiou} C. Anastasiou, S. Buehler, F. Herzog {\it et
  al.}, ``Total cross-section for Higgs boson hadroproduction with
  anomalous Standard Model interactions'', {\tt arXiv:1107.0683}. 
\bibitem{pasturmucci} G. Passarino, C. Sturm, S. Uccirati, ``Higgs 
  Pseudo-Observables, Second Riemann Sheet and All That'', {\it
    Nucl. Phys.} {\bf B834} (2010) 77-115, {\tt arXiv:1001.3360}. 

\end{thebibliography}

\end{document}